# Marked changes in electron transport through the blue copper protein azurin in the solid state upon deuteration

Nadav Amdursky[a,b], Israel Pecht[c], Mordechai Sheves[b,*], and David Cahen[a,*]

Departments of [a]Materials and Interfaces, [b]Organic Chemistry, and [c]Immunology, Weizmann Institute of Science, Rehovot 76100, Israel

* Corresponded authors:

Prof. David Cahen

Department of Materials and Interfaces, Weizmann Institute of Science, Rehovot 76100, Israel

Phone: +972-8-934-2246

e-mail: David.Cahen@weizmann.ac.il

Prof. Mordechai Sheves

Department of Organic Chemistry, Weizmann Institute of Science, Rehovot 76100, Israel

Phone: +972-8-934-4320

e-mail: mudi.sheves@weizmann.ac.il

**Abstract**

Measuring electron transport (ETp) across proteins in the solid-state offers a way to study electron transfer (ET) mechanism(s) that minimizes solvation effects on the process. Solid state ETp is sensitive to any static (conformational) or dynamic (vibrational) changes in the protein. Our macroscopic measurement technique extends the use of ETp measurements down to low temperatures and the concomitant lower current densities, because the larger area still yields measurable currents. Thus, we reported previously a surprising lack of temperature-dependence for ETp via the blue copper protein azurin (Az), from 80K till denaturation, while ETp via apo-(Cu-free) Az was found to be temperature dependent ≥200K. H/D substitution (deuteration) can provide a potentially powerful means to unravel factors that affect the ETp mechanism at a molecular level. Therefore, we measured and report here the kinetic deuterium isotope effect (KIE) on ETp through holo-Az as a function of temperature (30-340K). We find that deuteration has a striking effect in that it changes ETp from temperature independent to temperature dependent above 180K. This change is expressed in KIE values between 1.8 at 340K and 9.1 at ≤180K. These values are particularly remarkable in light of the previously reported inverse KIE on the ET in Az in solution. The high values that we obtain for the KIE on the ETp process across the protein monolayer are consistent with a transport mechanism that involves through-(H-containing)-bonds of the $\beta$-sheet structure of Az, likely those of amide groups.

**Introduction**

The blue copper protein, azurin (Az), functions as an electron carrier in the bacterial energy conversion system (1, 2). The intramolecular electron transfer (ET) reactions in Az in solution have been studied extensively over the last decades by a variety of methods (3-7). In particular, the ET process between the disulfide radical, RSSR$^-$ and the Cu(II) center, or between one of the histidines that was Ru-labeled and the Cu(II) center, has been studied by pulse radiolysis (3) or flash-quench (4, 5) techniques, respectively. More recently, measurements of electron transport (ETp) across Az in the solid-state have been made, mostly by scanning probe techniques (8, 9). We have previously reported results of macroscopic ETp measurements via Az, and showed that this process is *temperature-independent* down to 80K, i.e., with no measurable thermal activation barrier (10, 11). Several factors within the protein may affect ETp through it, such as its primary and secondary structure, and the protein H-bonding network, as noted by Gray and coworkers for ET in Az (2). To assess the impact of H-bonding on the ETp through Az, we examined the kinetic H/D isotope effect (KIE). Here we are especially interested in its temperature-dependence, because even small differences should be measurable as deviations from the remarkable temperature independence, exhibited by protium Az.

The deuterium KIE on the ET between the RSSR$^-$ and the Cu(II) center process of Az in solution has previously been studied as a function of temperature (12). Interestingly, it was found that the deuterium KIE is < 1 ($k_H / k_D = 0.7$ at 298K) and this "inverse" KIE was interpreted as the result of increased negative activation entropy for the ET reaction in the protein in H$_2$O, compared to that in D$_2$O. The "inverse" KIE was therefore explained as being a result of differences in the ET driving force, caused by Az undergoing a slightly larger thermal expansion in H$_2$O than in D$_2$O (12). The effect of the solvent on the KIE of the redox mechanism of Az was confirmed by Chi et al., (13). They measured the isotope effect on the electrochemical ET rate between the Cu ion of Az to a gold electrode through an alkanethiol bridge, and concluded that their result was mainly due to the effect of the solvent. Carrying out studies in the solid-state allows focusing more on the intramolecular ETp, by eliminating solvent effects that were shown to have a crucial ef-fect on the process (12, 13). Indeed, we find (as reported below) that the deuterium KIE on the ETp process is significantly different from that previously observed in solution.

**Results**

An earlier published protocol was used for Az deuteration (12) and the extent of the protein's deuteration was determined by NMR. Figure S1, in the supplementary materials, shows the 1D-NMR spectra of protium- and deuterium-labeled Az. Examination of the amide bond region (at 7.5-9.5 ppm) indicates that the majority of these groups were fully deuterated. Additional protons in the protein might have also undergone deuteration (such as those of the imidazole ring of histidine residues), but these are not clearly observed in the NMR spectra. Comparing the other parts of the NMR spectra shows strong similarity between the protonated and deuterated samples. This indicates that the protein's structural parameters did not change significantly upon deuteration. To further examine the structural similarity between the two samples, we used additional spectroscopic techniques (Fig. 1). The circular dichroism (CD) spectra (Fig. 1a) are identical, showing that the deuterated protein retains its secondary structure of mainly β-sheets, which is consistent with previous results (14). Furthermore, the UV-Vis absorption of the charge transfer band of the Cu(II) (625 nm) and the photoluminescence properties of the single tryptophan residue of Az (Fig. 1b and c, respectively), which are also sensitive to the protein conformation, are not affected by deuteration, in agreement with the CD results.

Our earlier described method was used to form monolayers of Az on conductive Si surfaces (10, 11). Both the deuterated and protium-labeled Az surfaces yielded layers with similar ellipsometry-derived *optical* thickness of 18 Å. This value is about half that expected for a monolayer. However, as we explained earlier, this is what is expected for a monolayer, if we consider both the > 90% coverage and the voids between the proteins, that are resolved by atomic force microscopy (AFM) topography (15). The AFM topography (Fig. 2a) indicated similar morphology and roughness (inset of Fig. 2a) of the protium and deuterium samples.

The current-voltage (*I-V*) characteristics of the surfaces were measured with an Hg drop as a second contact (in addition to the Si substrate contact), with a 0.2 mm$^2$ geomet-rical contact area. Use of such large area contact yields proportionally higher currents than, e.g., use of a nanoscopic area as with scanning probe techniques. The higher currents allow measuring some 7 orders of magnitude smaller current densities, a critical issue for low temperature measurements. The measurements are made without significant change in the force applied on the sample, in contrast to scanning probe and most other methods that measure samples under compressive or tensile stress.

Comparing the room temperature (298K) current-voltage, *I-V* curves between deuterated and protium-labeled Az shows that the current density through protium-labeled Az is higher than that through deuterated Az (Fig. 2b), yielding a KIE value of 2.4±0.4 (measured at -0.05 V). This value is in contrast to the < 1 value that was measured for the intra-molecular ET from the RSSR$^-$ radical to the Cu(II) for Az in solution (KIE=0.7) (12). In the present solid-state measurements only the tightly bound water molecules that help maintain the native conformation remain, essentially eliminating the solvation effects, but leaving the intramolecular ones. The relatively high observed KIE suggests that the hydrogen-bonding network of the *β*-sheet array that supports the protein's structure, and that is affected by deuteration, plays a role in the ETp through (i.e., the "conductivity" of) the protein, as will be discussed below.

Temperature-dependent ETp measurements were carried out between 30 and 340K. Figure 3a shows the current density through the protein (measured at -0.05 V) as a function of 1000/T. As we reported recently, ETp through protium Az is temperature-independent over a similar range of temperatures (10) (cf. Fig. 3a). Strikingly, we find for deuterated Az two regimes, a temperature-independent one (≤180K) and *a thermally activated one* (190-340K, Fig. 3b). The ETp temperature dependence of the deuterated protein leads to a gradually increasing KIE (Fig. 3c) from 1.8±0.3 at high temperature (340K) to 9.1±1.1 in the low temperature regime (≤180K), in sharp contrast to the temperature-independent ETp observed for protium Az. The change of the current density of the deuterated sample, as function of temperature, from a thermally-activated regime in high temperatures to a temperature-independent regime at low temperatures, qualitatively resembles the behavior of apo-Az (Fig. 3d) (10), as discussed further, below.

In order to verify that the observed effect is indeed a protein-related one, we conducted the same experiment as the latter, only with a thin Au pad as a top contact instead of the used Hg drop (Fig. S2). The Au pad was brought into contact by using the lift on float on method (16), and yields similar *I-V* characteristics as the Hg drop (11). Au pad as a top contact allows reaching higher temperatures than Hg drop, temperatures that are needed in order to observe the effect of the proteins' denaturation on the ETp magnitude. As we previously observed in protium-Az (10), a sharp irreversible decrease in the current density is observed at the temperature of 360K. By using the Au pad as a top contact we also observed a similar drop in current density of the deuterated-Az at the similar temperature range (Fig. S2), along with the same temperature-dependent behavior as observed with the Hg drop as a top contact, indicating that the observed ETp behavior is indeed related to the protein itself.

**Discussion**

The KIE for ET processes is mostly related to the barrier for electron tunneling, with values that usually range between 1-3 (17-19). Changes in KIE values may be due to two kinds of contributions, intramolecular effects, caused by changes in the physicochemical properties of the molecule, and solvation effects that correspond to the different intermolecular reorganization processes existing in $H_2O$ and $D_2O$ (17, 20-22). Though $H_2O$ and $D_2O$ share similar physical properties (23), the differences in solvation-related KIE on ET processes are due to a difference in the Marcus solvent reorganization energies of the molecules in question. This difference is related to differences in dielectric constant, *D*, and refractive index, *n*, of the two solvents, where the reorganization energy, $\lambda \propto n^{-2}$ ($\propto D^{-1}$) (17, 24). The solvent reorganization energy may limit the applicability of the non-adiabatic ET model (see later in eq. 1), a limitation that emphasizes the importance of the solvent in the ET process (25, 26). Because the differences in dielectric constants and refractive indices between $H_2O$ and $D_2O$ are rather small (27, 28), calculations predict that the different solvent properties of $H_2O$ and $D_2O$ should result in only a minor isotope effect of ET processes (17).

The intramolecular contributions to the KIE stem from a change in the vibrational modes in the deuterated molecule. Classical experimental and theoretical studies of KIE of ET processes were done on metallo-organic complexes, where the metal is coordinated to the deuterated ligand. In a theoretical analysis of KIE of ET, Buhks et al. (17, 29) concluded that the H/D substitution can yield two different contributions, i.e., changes in the skeletal vibrational frequencies caused upon deuteration, and the difference in the contributions of the vibrational modes to the Franck-Condon factors (17). The former mostly causes only a minor KIE effect on ET processes. We can see how the latter, major contribution to the ET rate constant, $k_{ET}$, comes about by the expression for $k_{ET}$ in the Marcus theory for ET (30):

$$k_{ET} = \frac{2\pi}{\hbar} H_{D-A}^2 (FC) \qquad (1)$$

where $H_{D-A}$ is the degree of electronic coupling between D (donor) and A (acceptor), which also depends on their separation distance, (FC) is the Franck-Condon factor and $\hbar$ is the reduced Planck's constant. In our solid-state measurements the D-A separation is replaced by the separation of the two electrodes between which the ETp is measured. Thus, any measured KIE is due to ETp through the whole protein. Assuming that the electronic coupling (between the protein and the electrode) is comparable between the protium and deuterated samples, due to the similarity in the junction configurations, we conclude that the FC factors are the main cause for the ETp KIE. According to Buhks et al. (17), if the FC factors are the main cause for the KIE of the ET process, the KIE is expected to markedly increase with decreasing temperature from high temperatures, up to saturation at low temperatures, as our measurements indeed indicate (Fig. 3c).

The thermally activated ETp of the deuterated sample at ≥180K (Fig. 3b, black squares) can be fitted to an Arrhenius equation ($\propto \exp(-E_a / k_B T)$), which yields an activation energy of $E_a = 85$ meV (8.2 kJ/mol). As we showed previously (10), the temperature-independent ETp of protium-Az (which corresponds to zero activation energy) is affected by removing the Cu ion (apo-Az). The outcome of those changes is a thermally-activated ETp (Fig. 3d), over a similar temperature range as we report here, with activation energy of 320 meV (33 kJ/mol). Although the calculated activation energy for the deuterated Az is much lower than that of apo-Az, it is remarkable, *because the deuterated protein still has the Cu redox centre.*

The secondary structure of Az is composed mainly from *β*-sheet structure, which is supported by the hydrogen-bonding network of the amide backbone. Thus, substitution of the amide bond proton by deuterium should affect markedly the protein's hydrogen-bonding network (2, 31, 32). Consequently, this substitution should also affect the Cu coordination sphere of Az. The Cu coordination sphere is very sensitive to the hydrogen bonds in its vicinity, as shown by Marshall et al.,(33) who altered the hydrogen bonds in the second coordination sphere of the Cu site, causing marked changes in the protein's reduction potential. In this context, Farver et al. (12) showed that the difference in the reduction potential between the deuterated- and protonated-Az is also temperature-dependent, with increasing difference with decreasing temperature. This difference in the reduction potential may well express a difference in flexibility of the protein, mainly around the Cu site (34-36). Changes in the protein flexibility have two outcomes (37): a) a structural aspect, i.e., as the protein is less flexible (equivalent to stiffer bonding) the reorganization energy for ET will be larger; b) a dynamic aspect, i.e., a less flexible protein has a shallower energy landscape as function of the (ET) reaction coordinate, which changes the ET activation energy. As a consequence, the difference in ETp as a function of temperature between protonated- and deuterated-Az (Fig. 3) may be a result of a change in the protein's flexibility, induced by the change in the H(D)-bonds vibrational modes at the Cu coordination sphere. Although we do not know which vibrational modes are most affected by the H/D exchange, we can speculate that a more flexible protein (i.e., the deuterated-Az) will have lower energy frequency modes, $\hbar\omega$, in comparison to a more rigid protein (the protonated-Az), as shown both experimentally (38) and theoretically (39) in other types of proteins. Thus, in the high-temperature regime of the deuterated Az sample, the lower energy frequency modes may cause a thermally-activated ETp process, according to Marcus theory when $\hbar\omega \leq 2k_BT$ (30), as indicated by our observations.

Gray, Onuchic and coworkers (2) have considered three types of structural elements constituting ET pathways in proteins: covalent bonds, hydrogen bonds and through-space jumps. Exploring different pathways of ET in Az by computational modeling, they pointed out that hydrogen bonds are as important as covalent links, and, in fact, can be the primary facilitator for electron tunneling within the protein. The importance of hydrogen bonds has also been demonstrated by conductance measurements of peptide monolayers, by either changing the length of the peptide (40) or by denaturation of the peptide structure (41). The unique approach that we used here allows us to experimentally investigate the role of intramolecular hydrogen bonds in the ETp process, while excluding any solvent effect, without changes in stress, and, thus, in the conformation of the studied molecule, and with very high sensitivity. Moreover, the used approach is highly sensitive to any conformational changes of the protein, as seen for example in the ETp process via bacteriorhodopsin (42), where we observed a change in the ETp magnitude as the protein undergoes minor structural change. The fact that we observed completely temperature-independent ETp process via protium-labeled azurin suggests that there is not even minor structural change of the protein as a function of temperature.

As noted above, the KIE of an intramolecular ET process is temperature-dependent, where the KIE value should increase as the temperature decreases, up to a constant value at low temperatures, e.g., < 150 K for cytochrome c oxidation (43) or < 60 K in metallo-organo complexes (17). Reaching such low temperatures with the common electrochemical approach for ET measurements of proteins (either with a macroscopic electrode or by nanoscale electrochemical scanning tunneling microscopy) is extremely difficult. The solid-state experimental procedure that we used here allows us to reach such low temperatures. Together with the high sensitivity (large currents) that we get from using macroscopic electrodes (instead of nanoscopic ones), makes it possible to observe the high intramolecular KIE values, which were theoretically expected, experimentally. The macroscopic scale samples and electrodes not only allow detecting current *densities* through the protein that are up to 7 orders of magnitude smaller than what can be measured by nanoscopic, scanning probe-based technique, but also yields instant spatial averaging of the current density through the proteins. Those advantages, which lack in other approaches, are what enable the reliable measurements of the low current-densities that flow at low temperatures (as in the case of e.g., the deuterated- and apo-Az). The observed KIE reflects the high sensitivity of the solid-state ETp measurements to any change in the protein flexibility, a change that may well be very hard to observe in aqueous environment, where solvent effects can mask it.

## Conclusion

The room temperature ETp rate via deuterated-Az is markedly lower than that via the protium-labeled one, and the KIE values increase considerably as the temperature decreases from 1.8 (at 340K) to 9.1 (at ≤160K). These solid-state measurements allow excluding the effect of the bulk $H_2O/D_2O$ solvent on the ETp rate, and help focus on the intramolecular differences within the protein. The relatively large KIE values underline the significant role that the amide groups, and consequently the H-bonding network of the mainly *β*-sheet structure of Az, have in the solid-state ETp through the protein, and support the through-(H-containing)-bond mechanism of ETp across proteins.

## Materials and Methods

**Protein preparation**. *Alcaligenes faecalis* Az was generously provided by RP Ambler, and its isolation and purification procedures were as previously reported (44). The protocol of Farver et al. (12) was used for the deuteration process: The protonated Az (~1.5 mg) in 20 mM phosphate buffer at pH7 was diluted in a deuterated buffer containing 20 mM $D_3PO_4$ (D, 99% in 85% D2O, Cambridge Isotope Laboratories, Inc.) and 100 mM NaOD (D, 99.5% in 40% D2O, Cambridge Isotope Laboratories, Inc.) in $D_2O$ (99.9%, Sigma-Aldrich) at ~pD7. The diluted solution was repeatedly (10 cycles) concentrated by Amicon® Ultracell – 4k (Millipore), re-diluted in the deuterated buffer and left over-night. The final solution concentration for both of the protonated and deuterated samples was ~0.1 mM.

**NMR measurement**. NMR samples contained ~1 mg Az dissolved in 90% $H_2O$ / 10% $D_2O$ for the protium-labeled Az or 100% $D_2O$ for the deuterium-labeled Az, in 20mM $NaH_2PO_4$ or $NaD_2PO_4$, pH=7.0. All NMR spectra were acquired on a Bruker AVIII800 NMR spectrometer equipped with TCI CryoProbe. One-dimensional $^1$H NMR spectra and two-dimensional homonuclear Hartmann-Hahn (HOHAAHA) spectra (45, 46) were acquired at 303K using excitation sculpting sequence (47) for water suppression. Mixing time used in the 2D-HOHAHA experiments was 80 ms. All spectra were processed and analyzed using Topspin (Bruker Biospin, DE) software.

**Spectroscopic analysis**. ChirascanT Circular Dichroism Spectrometer (Applied Photophysics) was used for the CD analysis. Cary 5000 UV-vis-NIR spectrophotometer (Varian, a part of Agilent Technologies, USA) was used for the optical absorption measurements. FluoroLog − Modular Spectrofluorometer (Horiba Jobin Yvon) was used for the photoluminescence measurements. A rectangular quartz cuvette with a pathlength of 3 mm was used for the CD and optical absorption measurements. A 5 mm diameter square quartz cuvette was used for the photoluminescence measurements.

**Surface preparation**. The protein monolayer surface was prepared as previously reported (11) with few modifications: Highly doped (<0.005 Ω cm) p-type Si surface <100> was cleaned by bath sonication in ethyl acetate/acetone/ethanol (2 min in each), followed by 30 min of piranha treatment (7/3 v/v of $H_2SO_4/H_2O_2$) at 80°C. The Si surface was then thoroughly rinsed in Milli-Q (18 MΩ) water, dipped in 2% HF solution for 1.5 min, to etch the Si surface (leaving a Si-H surface), put in fresh piranha for ~5 sec for controlled growth of thin oxide layer (9-10 Å), and the surface was then immediately rinsed thoroughly in water and dried under a nitrogen stream. A monolayer of 3-Mercaptopropyl trimethoxysilane (3-MPTMS, SH-terminated linker, 95%, Sigma-Aldrich) was prepared by immersing the $SiO_2$ substrate in 10 mM 3-MPTMS in bicyclohexyl for 1 hour, followed by 3 min of bath sonication in acetone and 10 sec in hot ethanol, yielding a monolayer thickness of ~7 Å. The 3-MPTMS coated surface was immersed in either the protonium labeled Az solution (~1 mg/ml) or the deuterated Az in the deuterated buffer for 2 hours, where the Az was covalently connected to the SH group of the linker via its cysteine (Cys26-Cys3) thiolate.

**AFM imaging**. The topography of the self-assembled monolayer of proteins was characterized by AFM in tapping mode. A Solver P47 SPM system (ND-MDT, Zelenograd Russia) and Si probes (NSC36, 75kHz, 0.6 N/m, MIKROMASCH) were used.

**Current-voltage measurements**. The back (InGa) and top (Hg) contacts were prepared and deposited as previously reported (11). The Hg was places on top of the protein monolayer by capillary. All the results were obtained on macroscopic areas (~0.2 mm$^2$ contact area, containing $10^9$-$10^{10}$ protein molecules). The sample was placed in low vacuum (0.1 mbar) chamber in a TTPX cryogenic system (Lakeshore), and both the sample holder and the probes were cooled. The temperature was monitored and controlled with an accuracy of 0.2 K. Between each change in temperature the sample was not measured, so as to allow it to reach thermal equilibrium.

**Acknowledgments**

We thank T. Scherf for assisting with the NMR measurements. We also thank H. Gray, M. Bixon and A. Nitzan for fruitful discussions. N.A thanks the Clore program for financial support. We thank the Minerva Foundation (Munich) for partial support. M.S holds the Katzir-Makineni chair in Chemistry. D.C holds the Schaefer Chair in Energy Research.

**Figures**

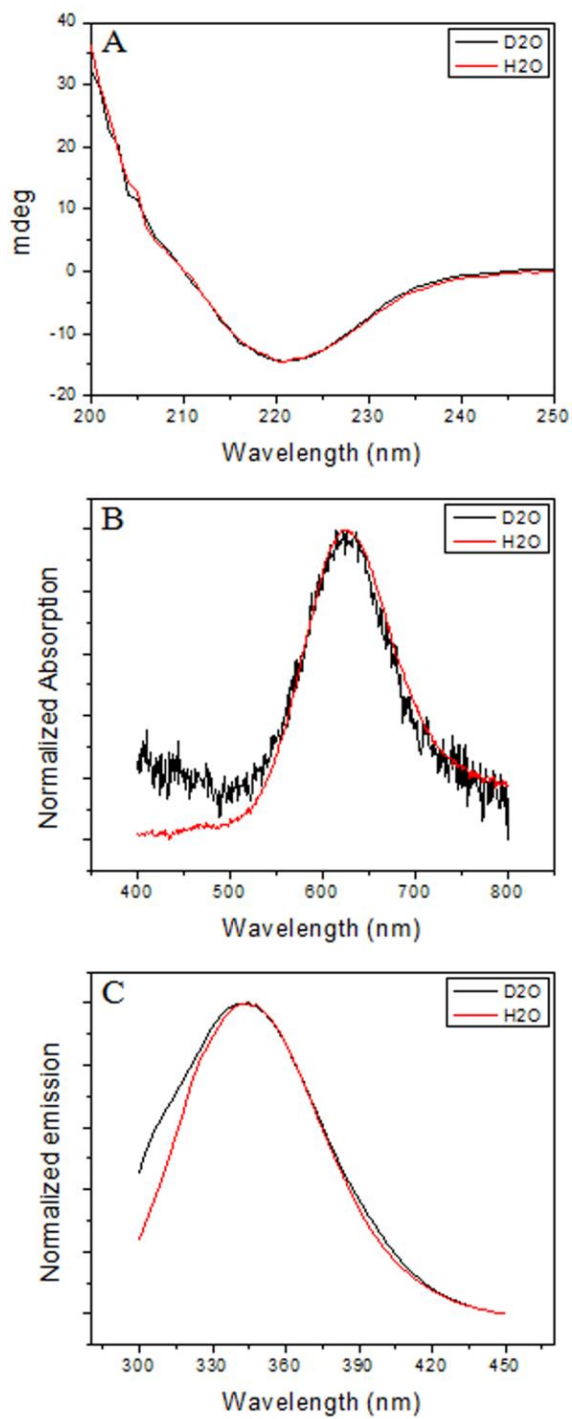

**Figure 1**. Spectroscopic analysis of the structure. (a) CD, (b) UV-Vis and (C) photoluminescence spectra of protium- and deuterium-labeled azurin (red and black curves, respectively).

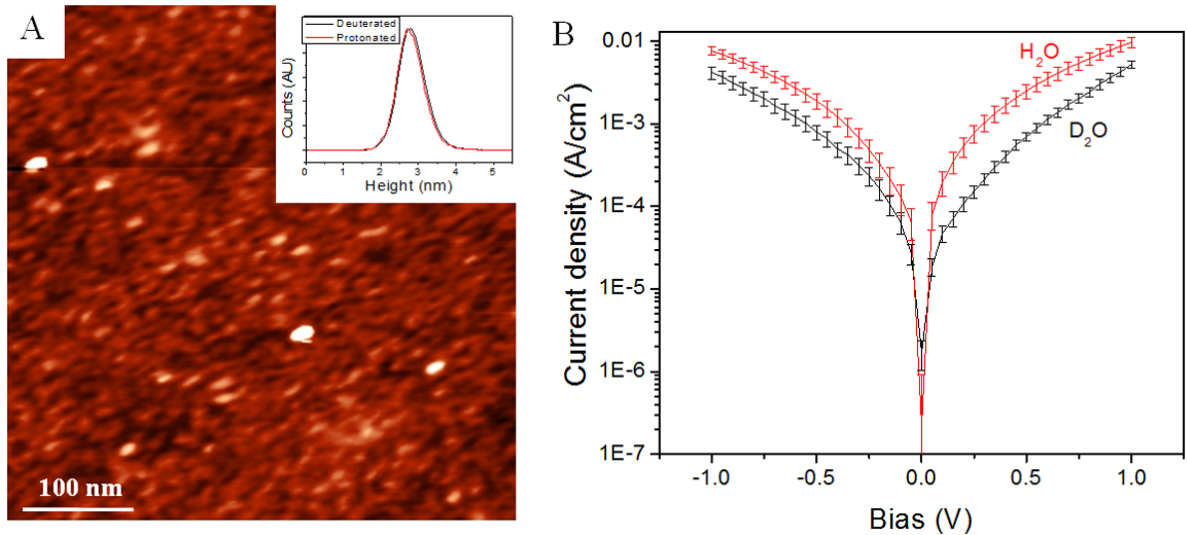

**Figure 2**. Surface morphology and current-voltage. (a) AFM image of Az coated surface. The z-scale is 7 nm for both of the images. The inset shows a superimposed histogram of the surface roughness of the protium and deuterated Az surfaces. (b) Current density vs. voltage at room temperature of protium- and deuterium-labeled Az surfaces.

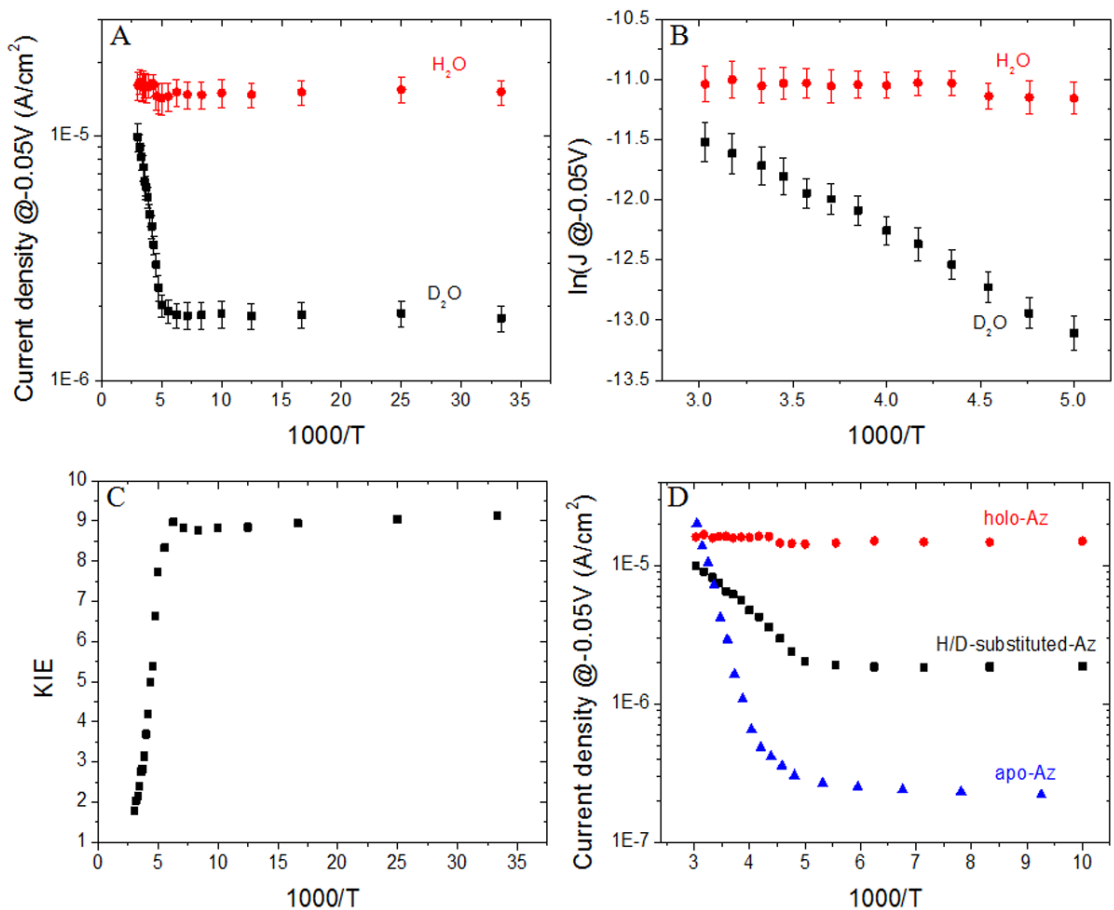

**Figure 3**. ETp behavior. (a) Current density at -50 mV *vs*. inverse temperature of protium- and deuterium-labeled Az. (b) zoom-in on the high temperature part of (b) on a *ln*(J) scale. (c) Temperature dependence of the deuterium KIE. (d) A comparison between the effects of H/D substitution to the removal of Cu-ion (apo-Az).

Nadav Amdursky[a,b], Israel Pecht[c], Mordechai Sheves[b,*], and David Cahen[a,*]

Departments of [a]Materials and Interfaces, [b]Organic Chemistry, and [c]Immunology, Weizmann Institute of Science, Rehovot 76100, Israel

**Supplementary Figure legends**

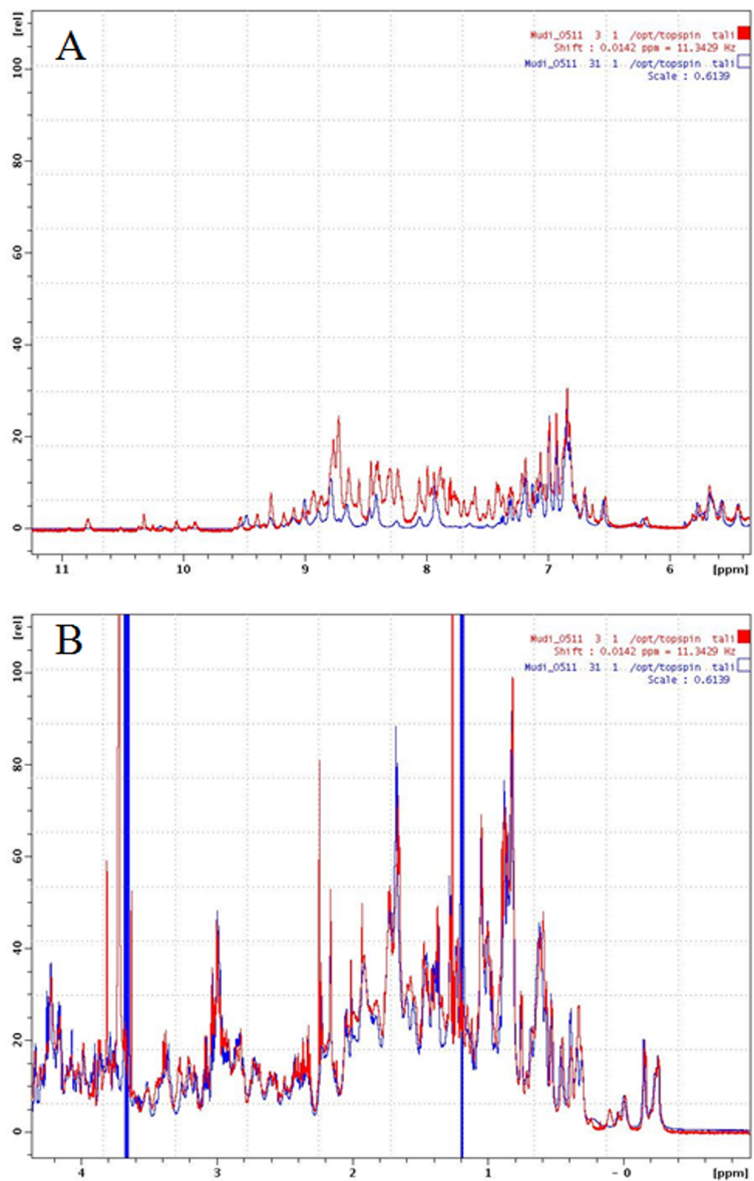

**Figure S1**. Superposition of one-dimensional 1H NMR spectra of protium- and deuterium-labeled azurin (red and blue curves, respectively). (a) Aromatic and amide region of the spectrum, showing signals of aromatic and exchangeable amide protons. (b) Aliphatic region of the spectrum, showing signals of non-exchangeable protons.

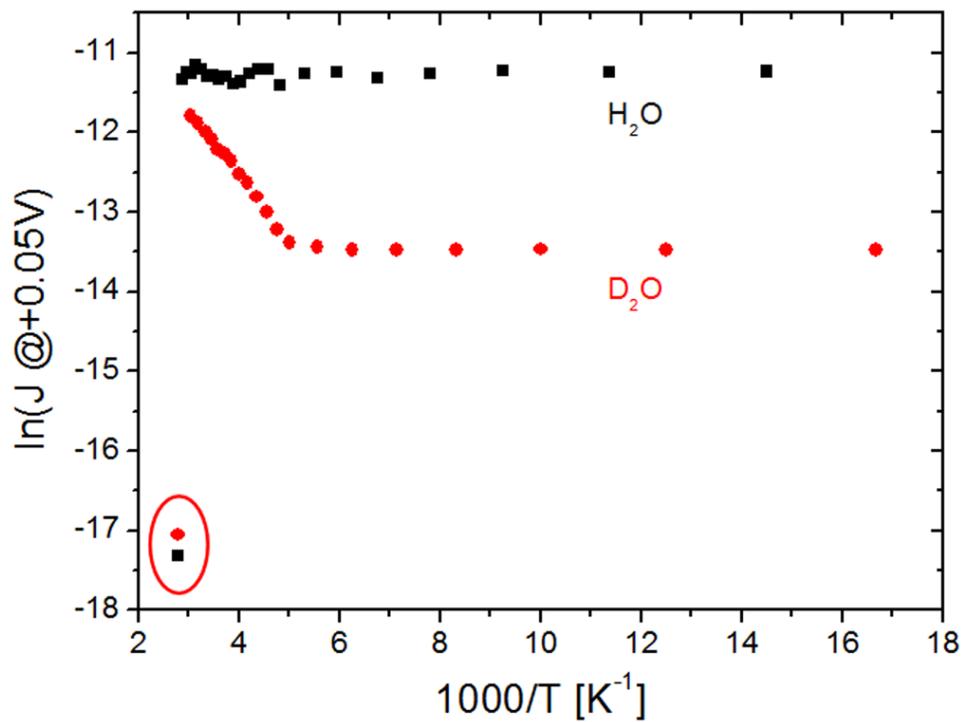

**Figure S2**. Current density at -50 mV *vs*. inverse temperature of protium- and deuterium-labeled Az with a Au pad as top contact (i.e., by LOFO). The red circle indicates the denaturation point of the protein.